\documentclass[12pt,preprint]{aastex}
\def\uka { \raisebox{-0.5ex} {\mbox{$\stackrel{<}{\scriptstyle \sim}$}}}
\def\uga { \raisebox{-0.5ex} {\mbox{$\stackrel{>}{\scriptstyle \sim}$}}}

\def\next{\,\!}
\slugcomment{The Astrophysical Journal, in press}
\shortauthors{H. Krawczynski}
\shorttitle{X/TeV Emission from Parallel Beams}
\begin{document}
\title{X-ray and TeV Gamma-Ray Emission from Parallel Electron-Positron or Electron-Proton Beams in BL~Lac~Objects}
\author{H.~Krawczynski}
\affil{Washington University in St. Louis, Physics Department,\\ 
1 Brookings Drive, CB 1105, St. Louis, MO 63130}
\email{krawcz@wuphys.wustl.edu}
\begin{abstract}
In this paper we discuss models of the X-rays and TeV $\gamma$-ray 
emission from BL Lac objects based on parallel electron-positron 
or electron-proton beams that form close to the central black hole
owing to the strong electric fields generated by the accretion disk 
and possibly also by the black hole itself. 
Fitting the energy spectrum of the BL Lac object Mrk 501, we obtain
tight constraints on the beam properties.
Launching a sufficiently energetic beam requires rather strong magnetic 
fields close to the black hole ($\sim 100-1000$ G). However, the model 
fits imply that the magnetic field in the emission region is 
only $\sim 0.02$ G. Thus, the particles are accelerated close to the 
black hole and propagate a considerable distance before instabilities 
trigger the dissipation of energy through synchrotron and 
self-Compton emission. We discuss various approaches to generate 
enough power to drive the jet and, at the same time, to 
accelerate particles to $\sim$20~TeV energies.
Although the parallel beam model has its own problems, it explains 
some of the long-standing problems that plague models based on 
Fermi-type particle acceleration, like the presence of a 
very high minimum Lorentz factor of accelerated particles. 
We conclude with a brief discussion of the implications of the model 
for the difference between the processes of jet formation in 
BL Lac type objects and in quasars.
\end{abstract}
\keywords{galaxies: jets --- galaxies: BL Lacertae objects: 
individual (Mrk 501) --- gamma rays: theory --- X-rays: galaxies}
\section{Introduction}
\subsection{Observations and models of the continuum emission from blazars}
\label{intro}
Observations with the {\it EGRET Energetic Gamma-Ray Experiment Telescope}
on board of the Compton Gamma-Ray Observatory (CGRO) revealed that
blazars are powerful and variable emitters, not just at radio through
optical wavelengths but also at $\ge$100 MeV $\gamma$-ray 
energies \citep{Hartman1999}. 
The 66 blazars detected with EGRET were mainly quasars, i.e. Flat Spectrum Radio Quasars and Optically 
Violent Variables. Observations with ground based Cerenkov telescopes showed that BL Lac objects, 
the low-power counterparts to the high-powered EGRET quasars, emit even more energetic 
$\gamma$-rays \citep{Punch1992}. 
Based on observations with ground based Cerenkov telescopes, more than
a dozen BL Lac type objects have been identified as sources of $>$300
GeV gamma-rays \citep{Aharonian2005,Horan2004}.

The MeV and TeV $\gamma$-ray emission from blazars is commonly thought 
(e.g.\ \citealt*{Tavecchio2005,Krawczynski2006})
to originate from relativistic particle dominated outflows (jets) from mass accreting supermassive
black holes \citep{Lynden-Bell1969,Zeldovich1971,Rees1978}. 
The jet may form electromagnetically or through magnetohydrodynamic processes.
Electromagnetic models come in two flavors. Either the accretion 
disk \citep{Lovelace1976,Blandford1976} or the Kerr 
black hole \citep{Blandford1977} launch a Poynting flux 
dominated flow. The mechanism to convert the Poynting dominated outflow
into a particle dominated one is not  understood. In other models 
magnetohydrodynamic pressure may play a dominant role in the process 
of jet formation \citep{Blandford1982,Begelman1984}.
Models usually assume that the particle dominated outflows move with velocities $v\,\sim$ $c$ 
and Bulk Lorentz factors $\Gamma\,=$ $(1-(v/c)^2)^{-1/2}$ between a few and $\sim$50.
Some models suggest very high bulk Lorentz factors between 
100 and 1000 \citep{Pohl2000,Rebillot2006}. 
At shocks within the jet plasma, the Fermi mechanism may transfer a fraction of the bulk 
kinetic energy into random kinetic energy of high-energy particles, electrons, 
or protons \citep{Rees1978}. These accelerated particles themselves, or secondaries 
produced in cascades, may emit the observed continuum emission. 
Although the spectral energy distributions (SEDs) of blazars are often only 
sparsely sampled, most blazars seem to emit two distinct emission 
components, one at low energies and one at high energies. 
The two components are attributed either to the emission by a single particle 
population emitting photons of vastly different energies through two different emission processes 
\citep{Rees1967,Blandford1978,Konigl1981,Ghisellini1985}, or to the emission 
by two particle populations. An example for the first are synchrotron Compton models in 
which a single population of electrons emit the low-energy and high-energy 
emission components as synchrotron and inverse Compton emission, respectively.
In synchrotron self-Compton (SSC) models, the synchrotron photons are the dominant 
source of target photons for inverse Compton processes.
Examples for the latter are hadronic models in which the 
low-energy component originates as synchrotron radiation from a population of 
low-energy electrons. The high-energy component is synchrotron emission either 
from extremely high energy (EHE) protons \citep{Aharonian2000,Muecke2001,Muecke2003}, 
or from secondary $e^+/e^-$ resulting from a synchrotron and pair-creation cascade 
initiated by EHE protons \citep{Mannheim1993} or high-energy electrons or photons 
\citep{Lovelace1979,Burns1982,Blandford1995,Levinson1995}.

Following the end of the CGRO in 2000, studies of the MeV emission from quasars 
has to await the launch of the next space-borne $\gamma$-ray telescope. The  {\it GLAST} 
{\it (Gamma-Ray Large Space Telescope)} satellite will be one order of magnitude 
more sensitive than {\it EGRET} and is scheduled for launch in the near future.
Ground-based $\gamma$-ray observatories continue to provide data on the 
TeV $\gamma$-ray emission from BL Lac objects.
Numerous broadband multiwavelength observation campaigns on a few objects
(most notably Markarian (Mrk) 421 ($z\,=$ 0.031), Mrk 501 ($z\,=$ 0.034), 
and 1ES 1959+650 ($z\,=$ 0.048)) have yielded important results which shed 
some light on the emission mechanism.
The key-results from these campaigns are: 
\begin{enumerate}
\item There is a highly significant flux correlation between X-rays and 
TeV $\gamma$-rays \citep{Takahashi1996,Buckley1996,Krawczynski2000,Sambruna2000,Fossati2004}. The ``time lag'' 
between the flux variations at X-rays and at TeV $\gamma$-rays is sufficiently 
short (\uka 1 hr) to evade detection \citep{Maraschi1999,Fossati2004}.
The X-rays and TeV $\gamma$-rays are emitted close to the peaks of the
low-energy and high-energy emission components, respectively.
\item The sources show strong correlated X-ray/TeV $\gamma$-ray flares with X-ray and 
TeV $\gamma$-ray flux changes by factors of between a few and 20 on time 
scales between 15 min (Mrk 421) and a few hours (Mrk 501, 1ES 1959+650) 
\citep{Gaidos1996,Krawczynski2000}.
\item The campaigns did not reveal a highly significant flux correlation 
of the radio to optical emission with the X-ray or TeV $\gamma$-ray emissions. 
The interested reader is referred to \citet{Buckley1996} for weak evidence for
such a correlation and to \citet{Blaz2005,Rebillot2006} 
for campaigns which did not reveal supportive evidence. 
\item As discussed further below, shock acceleration theories predict
that the X-ray and TeV $\gamma$-ray energy spectra are softer during 
the early rising phases of flares than during the later phases. 
While some flares showed such a behavior, others did 
not (e.g.\ \citealt*{Takahashi2000,Falcone2004,Fossati2004}). 
\end{enumerate}
Despite some disparity of the data on a whole, the good X-ray/TeV-$\gamma$-ray 
correlation strengthens the case for leptonic synchrotron-Compton models.
\subsection{The reference model and its problems}
In this paper, we focus the discussion on the X-ray and TeV-$\gamma$-ray emission 
from BL Lac type objects. In Section \ref{discussion}, we will briefly outline
the relevance of the model for quasars. 

We use the term ``reference model'' for the following combination of model components. 
The accretion system launches a Poynting flux dominated jet and as the outflow propagates, 
the flow transforms into a particle dominated outflow. Shocks within the jet transfer 
a fraction of the jet's bulk kinetic energy to a few high-energy 
particles that emit synchrotron and Inverse Compton emission.

The reference model suffers from a number of weak links. No simple mechanism 
has yet been suggested to convert the Poynting flux dominated outflow into 
a particle dominated outflow. Furthermore, the hypothesis of shock 
acceleration of electrons to $\sim$TeV energies is observationally very 
poorly supported. The modeling of the SEDs of some BL Lacs require 
``non-standard'' electron energy spectra.

Models of Mrk 501 and Mrk 421 data require a minimum Lorentz factor of 
accelerated particles $\gamma_{\rm min}$ on the order of $10^5$ in the 
jet reference frame (e.g.\ \citealt*{Pian1998,Krawczynski2002}), or 
non-power-law distributions with very high characteristic Lorentz factors 
(e.g.\ \citealt*{Sauge2004,Katarzynski2006,Giebels2006}).
If the shocks are internal to the jet and the jet medium is made of 
cold protons, simple arguments lead to $\gamma_{\rm min}$-values 
close to the proton-to-electron mass ratio  
$\gamma_{\rm min}\,\approx$ $m_{\rm p} / m_{\rm e}\,=$ 1836.
If the shocks are external and the jet medium runs into a much 
slower target medium, 
the $\gamma_{\rm min}$-values could be higher by a factor of $\Gamma$. 
A modification of the reference model might be able to account for the 
high $\gamma_{\rm min}$-values and for the non-standard electron
spectral indices even in the internal shock model. \citet{Bykov1996} 
point out that statistical acceleration by relativistic magnetohydrodynamic 
fluctuations in flow collision regions of jets might give rise to hard electron 
energy spectra and to $\gamma_{\min}$-values on the order of 10$^5$. 

As mentioned above, the theory of shock acceleration predicts that the X-ray and TeV $\gamma$-ray
spectra of flares are soft during the early rising phases of flares.
The beautiful Mrk 421 observations taken in 2001 with the {\it RXTE (Rossi X-ray Timing Explorer)} 
show both, soft and hard energy spectra during the early rising phases of flares
(\citealt*{Fossati2004}, and Fossati and Buckley, private communication, 2006). 
This negative result can still be explained in the framework of the reference model,
if a strong flare is made of the superposition of many small flares, or if
other processes  (e.g.\ the growth and decline of the shock, a changing viewing angle,
radiative cooling, and adiabatic particle losses) dominate the temporal evolution 
of the flares \citep{Kirk1999}.
\citet{Sokolov2004} and \citet{Sokolov2005} emphasize that the geometry 
and structure of the acceleration region also influence the observed spectral behavior.

A different problem concerns the bulk Lorentz factors of the emitting 
jet plasmas. Simple SSC models require bulk Lorentz 
factors $\Gamma\,\uga$25 \citep{Krawczynski2001,Konopelko2003,Henri2006}. 
While Very Large Baseline Array (VLBA) observations of quasars have recently succeeded 
in finding sources with apparent superluminal motions that support $\Gamma\,\sim$ 50 \citep{Piner2006}, the 
observations of BL Lac objects in general \citep{Lister2006} and the TeV sources 
Mrk 421, Mrk 501, 1ES 1959+650 in particular \citep{Piner2004,Piner2005} show only 
subluminally moving components. 
In contrast to SSC models, ``External Compton'' models do not require such high bulk 
Lorentz factors. In these latter models an otherwise unobserved radiation component 
originating from a different emission zone than the X-ray and TeV $\gamma$-ray emission, 
provides the seed photons for the inverse Compton processes that produce 
the observed $\gamma$-rays. The recent models of \citet{Ghisellini2005} and 
\citet{Georganopoulos2003} can fit the data with lower bulk Lorentz factors.
\subsection{Structure of the paper}
Motivated by the problems of the reference model, we explore here models 
similar to the ones of \citet{Lovelace1976}, \citet{Blandford1976}, 
and \citet{Blandford1977}. As the models invoke parallel electron-positron or electron-proton
beams, we will refer to them in the following as ``parallel beam'' models. 
A disk carrying a magnetic field and possibly the Kerr black hole induce a strong 
electric field parallel to the rotation axis of the black hole/disk system. 
We assume that the electric field will give rise to a beam of high-energy electrons 
and positrons or protons that move nearly parallel to the rotation axis of the accretion system.
These particles, in turn, directly emit the observed X-ray and TeV $\gamma$-ray emission as synchrotron and inverse 
Compton emission, respectively. This model does not require the formation of shocks 
or Fermi-type acceleration mechanisms as in the reference model. 
In this paper, we scrutinize for the first time how the model can be applied to 
explain the X-ray/TeV $\gamma$-ray data that have recently become available.
We will use the data to guide the evaluation of the models. 
In Section \ref{beam}, we will first infer the properties of the nearly 
parallel beam from modeling a specific data set. 
Once that we know the properties of the beam, we will discuss 
in Sect.\ \ref{BeamFormation} where and how the accretion system may form such a beam. 
Finally, in Sect.\ \ref{discussion} we will summarize the findings 
and discuss the strengths and weaknesses of the model.

Electromagnetic models have recently been discussed by \citet{Levinson2000}, 
\citet{Maraschi2003}, \cite{Kundt2004}, and \citet{Katz2006}.
The work described in the following is new in that it starts with 
simultaneously modeling the X-ray {\it and} TeV $\gamma$-ray data.
The data constrain the beam properties tightly, and it becomes possible
to perform a targeted study of how and where the beam originates.
We will describe the parallel beams in the AGN frame and we will focus 
on electron and positrons as emitting particles. To make the description 
more concise, we will frequently use the term electrons for both, electrons and positrons.
The best studied BL Lac objects Mrk 421, Mrk 501, and 1ES 1959+650 have all 
redshifts well below 0.1, and we will omit all redshift and $k$-correction 
factors to avoid unnecessary clutter in the equations.
\section{Beam Properties}
\label{beam}
\subsection{Model Geometries}
We will model the X-ray and TeV $\gamma$-ray observations of Mrk~501 taken on April 16th, 
1997, the day of the strongest flare, observed during a spectacular six months long flaring period.
Over the six month period several X-ray \citep{Pian1998,Catanese1997,Krawczynski2000} and 
TeV $\gamma$-ray observatories \citep{Aharonian1999,Djannati-Atai1999,Quinn1999} gathered good data. 
We will use the data from the {\it BeppoSAX (Satellite per Astronomia X)} and {\it CAT (Cerenkov 
Array at Themis)} experiments that observed the April 16 flare with partial temporal overlap.
The Mrk 501 data taken in 1997 have been modeled by many different 
groups (e.g.\ \citealt*{Pian1998,Krawczynski2002,Mannheim1998,Muecke2001}).
We consider two model geometries of the emission zones to get an estimate of the model 
dependencies of the relevant quantities. In both cases, we assume a magnetic field 
configuration with azimuthal symmetry with regards to the jet axis. 
The charged particles move along the magnetic 
field lines that are nearly parallel to the jet axis (the $z$-axis). 
At the outer edges of the emission zone, the field lines make an 
angle $\alpha_{\rm max}$ to the jet axis.

\begin{figure}[bh]
\begin{center}
\resizebox{16cm}{!}{\plotone{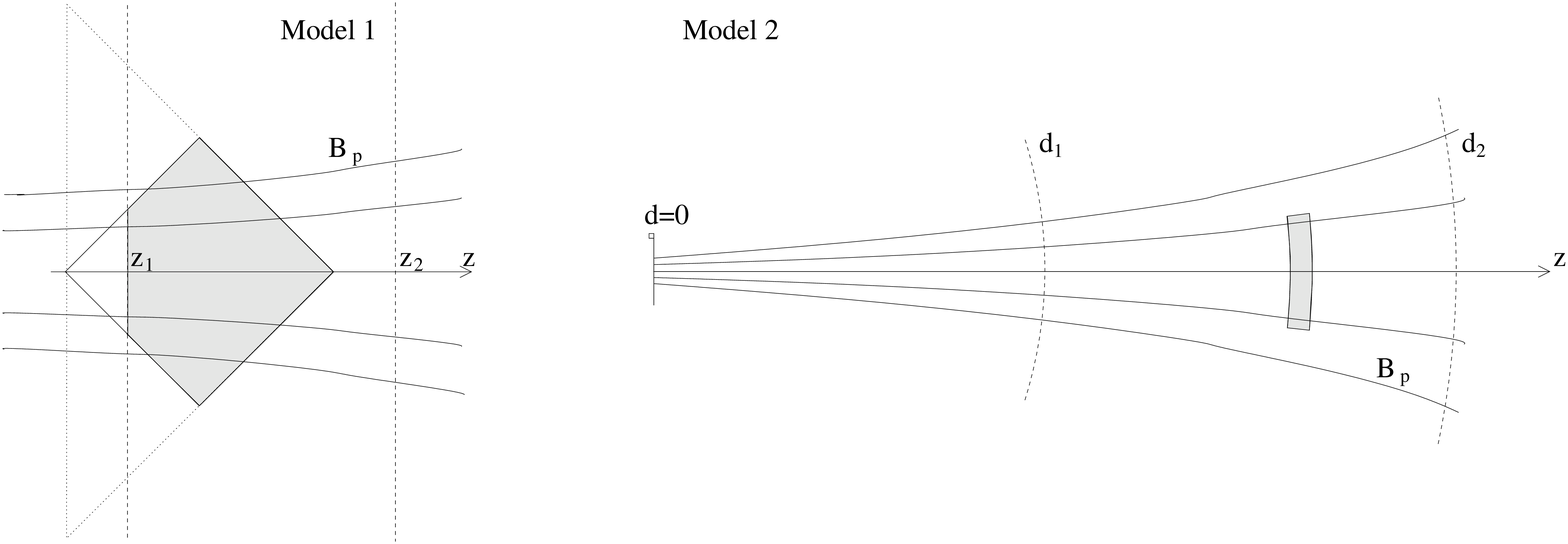}}
\end{center}
\caption{\label{CONE} Two geometries to account for the X-ray and TeV $\gamma$-ray emission
from BL Lac objects. On the left side, the emitting particles fill a volume (shaded grey here) 
made up of two back-to-back cones. The geometry might result from a single larger cone (dotted 
line), where the number of electrons tapers off after some time. The right side shows a shell of 
high-energy particles traveling down the jet. Such a geometry might result
from an electromagnetic shower developing close to the black hole. 
In both geometries, the emitting electrons and/or positrons drift 
along the magnetic field lines, and start emitting within the regions delimited
by $z_1$ and $z_2$, and by $d_1$ and $d_2$, respectively. The true geometry might 
lie between the two extremes shown here.
}
\end{figure}

In Model 1 the high-energy electrons fill a shape that resembles two back-to-back cones with 
a maximum cone radius of $R$ and a total length of $2\,R$ along the jet axis 
(Fig.\ \ref{CONE}, left side). 
The specific geometry chosen here reproduces the triangle shaped flares 
commonly observed in X-ray and TeV-$\gamma$-ray light curves of blazars.
Note that the exact shape of the region is not very important. 
The important property of Model 1 is that the diameter divided by the speed 
of light equals approximately the observed flare duration. For simplicity we 
choose an identical height and length of the emitting volume. More realistic shapes 
may be obtained from modeling how an $e^+/e^-$ spark forms in a high electric field region.
We assume that the leptons follow the magnetic field lines dissipating little energy 
until they enter the radiation zone at a distance $z_1$ from the black hole 
in which some instability causes the particles to move with an isotropic 
distribution of pitch angles $\theta$ to the magnetic field lines with $\theta<\theta_{\rm max}$. 
We assume that the electrons stop emitting when they exit the radiation 
zone at a distance $z_2$ from the black hole with $z_2-z_1\,=$ $2\,R$. 
The electrons may stop emitting because further instabilities slow them down.
The duration $\Delta t_{\rm obs}$ of a flare is given by the length of the 
emission zone: $\Delta t_{\rm obs}\approx$ $(z_2-z_1)\,/\,c\,=$ $2\,R\,/\,c$.
For Mrk 501, typical flare durations are $\Delta t_{\rm obs}=$ 12~hrs and thus 
$R\,\approx$ 6.5 $(\Delta t_{\rm obs}/12{\rm ~hrs})$  10$^{14}\,$cm. 
Each electron spends a time $\Delta t_{\rm rad}=$ $\Delta t_{\rm obs}$ in the 
emission region. Here and in the following model, 
the emitting particles travel together with the emitted photons down the jet 
with the main velocity dispersion originating from the pitch angle 
distribution of the leptons and photons.

Model 2 describes a spatial electron/positron distribution that might 
result from a synchrotron/pair-creation cascade of extremely high-energy particles
accelerated by strong electric fields close to the black hole. 
We assume that the electrons move along magnetic field lines 
that make angles $\alpha$ up to $\alpha_{\rm max}$ to the jet axis and are 
concentrated in a spherical or conical shell or ``shower front'' 
that travels down the jet axis (Fig.\ \ref{CONE}, 
right side). We assume that the electrons only emit in a radiation 
zone that extends from a distance $d_1$ to $d_2$ from the black hole.
Within the emission zone the electrons spiral with pitch angles 
$\theta<\theta_{\rm max}$ around magnetic field lines.

We assume $d_1\equiv d_2/2$ and consider only the case where the jet points 
exactly at the observer. High-Lorentz factor electrons emit synchrotron and 
inverse Compton emission only along the direction of their motion.
Taking into account the electron pitch angle distribution, 
observers can only see the emission from electrons traveling 
along field lines with $\alpha\,\le$ $\theta_{\rm max}$.
Based on the standard equations from the description of superluminal 
motion \citep{Blandford1977b} and neglecting 
$(\cos{(\theta_{\rm max})})$ factors, 
we get
\begin{equation}
\Delta t_{\rm obs}\,\approx\,\frac{\kappa\, d_2\,+\,T}{c} {\rm ~~,~with~~}
\label{d2}
\end{equation}
\begin{equation}
\kappa\, = \, 1\,-\,\cos{\theta_{\rm max}} \,
\approx\,\frac{1}{1642}\,\left(\frac{\theta_{\rm max}}{2^{\circ}}\right)^2
\end{equation}
As the electrons travel from $d_1$ to $d_2$, their different pitch angles 
result in an increase of the thickness of the
shell by $\Delta T\,\approx$ $\kappa\,(d_2-d_1)$ . 
Taking into account that the shell has 
already a finite thickness at $d_1$, we use in simple approximation a constant 
shell thickness of
\begin{equation}
T\,=\,\kappa\,d_2
\label{T}
\end{equation}
Combining eqs.\ (\ref{d2}) and (\ref{T}), we get
\begin{equation}
d_2\,\approx\,\frac{c\,\Delta t_{\rm obs}}{2\,\kappa}\,\approx\,
10^{18}~
\left(\frac{\Delta t}{\rm 12~hrs}\right)
\left(\frac{\theta_{\rm max}}{2^{\circ}}\right)^{-2}~\rm cm
\end{equation}
The thickness is $T\,=$ $c\,\Delta t_{\rm obs}\,/2\,=$ 
6.5$\cdot 10^{14}$ $(\Delta t_{\rm obs}/{\rm 12~hrs})$~cm.
At $d_2$ the shell has the following height perpendicular to the jet axis
\begin{equation}
H\,=\,\frac{c\,\Delta t_{\rm obs}\,\sin{\theta_{\rm max}}}{2 \kappa}\,\approx\,
3.7  \cdot 10^{16}~
\left(\frac{\Delta t}{\rm 12~hrs}\right)
\left(\frac{\theta_{\rm max}}{2^{\circ}}\right)^{-1}
~\rm cm
%
\end{equation}
For $d_1\,=$ $d_2/2$, each electron spends a time $\Delta t_{\rm rad}\,=$
$(d_2-d_1)/c\,=$ $\Delta t_{\rm obs}\,/\,(4\,\kappa)$ in the emission region.
In Model 2, triangle shaped pulses could result from a particle density that 
drops at a certain distance from the jet axis.

Models 1 and 2 represent extreme geometries. The true geometry may lie 
between these two extremes.
\subsection{Simple Analytical Considerations}
In this section, we use some simple analytical estimates to derive insights 
about the beam parameters and their scaling behavior with the model parameters.
Further below, we will see that the inverse Compton emission is mainly emitted
in the Klein-Nishina regime where the electrons give most of their energy
to the scattered photons.
The observation of $\sim$20 TeV photons thus implies the presence of electrons 
with Lorentz factors $\gamma_{  0}\,=$ 4$\cdot 10^7$ and energy 
$E_{\rm e}\,=$ $\gamma_{  0} m_{\rm e} c^2\,=$ 20~TeV. 
From the {\it BeppoSAX} X-ray observations on April 16, 1997 \citep{Pian1998,Massaro2004} 
we infer an 0.1 keV-200 keV energy flux of 
$\hat{I}_{\rm x}\,=$ $4\cdot 10^{-9}$ ergs cm$^{-2}$ sec$^{-1}$.
The observations did not cover the entire flare, and we assume 
here that the flux averaged over the flare duration 
was half the peak flux $I_{\rm x}\,=\hat{I}_{\rm x}/2$.
If the source at distance $D$ emits into the solid angle 
$\Delta \Omega\,\approx$ $2\,\pi$ $(1-\cos{\sqrt{\alpha_{\rm max}\next ^2+\theta_{\rm max}\next ^2}})$, 
the time averaged X-ray luminosity during the flare is
\begin{equation}
L_{\rm x}\,=\,I_{\rm x}\,\Delta\Omega\,D^2 \,
\approx\,3 \cdot 10^{42}\,
{\rm~erg~sec}^{-1}
\end{equation} 
The total energy emitted into the X-ray band is 
\begin{equation}
W_{\rm x}\,=\,\Delta t_{\rm obs} \, L_{\rm x}
\end{equation}
Electrons with Lorentz factor $\gamma$ moving with pitch 
angles $0<\theta<\theta_{\rm max}\,\ll$ 1 emit synchrotron radiation 
at a mean critical frequency  
\begin{equation}
\nu_{\rm c}\,=\,\frac{e\,\gamma^2\,B\,\theta_{\rm max}}
{2\,\pi\,m_{\rm e}\,c}
\label{nuc}
\end{equation}
For a Lorentz factor $\gamma_{  0}$, the frequency of synchrotron 
emission equals $\nu_{\rm  s0}\,\equiv$ 2.4$\cdot 10^{17}$ Hz for a magnetic field
\begin{equation}
B\,= \frac{2\,\pi\,m_{\rm e}\,c\,\nu_{\rm  s0}}
{e\,\gamma_{  0}\next ^2\, \theta_{\rm max}}
\,\approx\,
0.016~
\left(\frac{\nu_{\rm  s0}}{2.4\cdot 10^{17}\rm~Hz}\right)
\left(\frac{\gamma_{\rm  0}}{4\cdot 10^{7}}\right)^{-2}
\left(\frac{\theta_{\rm  max}}{2^{\circ}}\right)^{-1}~\rm G
\label{B}
\end{equation} 
where $h$ is Planck's constant. Averaged over pitch angles, the 
synchrotron power per electron is
\begin{equation}
p_{\rm s} \,=\, \frac{e^4\,\gamma_{  0}\next^2 \, B^2\,\theta_{\rm max}\next ^2}{3\, m_{\rm e}\next^2\,c^3}
\,\approx\,3.9 \cdot 10^{-7}~
\left(\frac{\nu_{\rm  s0}}{2.4\cdot 10^{17}\rm~Hz}\right)^2
\left(\frac{\gamma_{\rm  0}}{4\cdot 10^{7}}\right)^{-2}
{\rm~erg~sec}^{-1}
\label{ps}
\end{equation}
The synchrotron cooling time $t_{\rm s}\,=$ $E_{\rm e}/p_{\rm s}$ is
\begin{equation}
t_{\rm s}\,=\frac{3\,m_{\rm e}\next^3\,c^5}{e^4\,\gamma_{\rm  0}\,B^2\,\theta_{\rm max}\next^2}\,\approx\,
8.3\cdot 10^7
\left(\frac{\nu_{\rm  s0}}{2.4\cdot 10^{17}\rm~Hz}\right)^{-2}
\left(\frac{\gamma_{\rm  0}}{4\cdot 10^{7}}\right)^{3}
{\rm~sec} 
\label{ts}
\end{equation}
In Model 1, the electrons spend a time $\Delta t_{\rm obs}$ in the emission zone and radiate
only $\sim t_{\rm s}\,/$ $\Delta t_{\rm obs}\,$ $\approx 0.05$\% of their energy 
before leaving it. Given that the total synchrotron and inverse Compton
luminosities are approximately equal, we see that the emitting particles  radiate away 
only 0.1\% of their energy. Model 1 is thus radiatively very inefficient.
The efficiency does not depend strongly on details of the shape of the emitting 
region, as long as the length of the emission region equals approximately 
the flare duration. The scaling with the model input 
parameters shows that the efficiency is independent of the electron pitch 
angle distribution. It is lower if electrons with energies 
exceeding 20~TeV emit the observed synchrotron emission.

In Model 2, the emission zone is more extended. For this model, a rough 
estimate shows that the combined synchrotron and inverse Compton cooling time 
$\left(t_{\rm s}\next^{-1}+t_{\rm ic}\next^{-1}\right)^{-1}$ equals approximately 
the time $\frac{1}{4\,\kappa}\,\Delta t_{\rm obs}$ that the emitting 
particles stay in the emission region. Model 2 is thus radiatively very efficient.

A total number of electrons 
\begin{eqnarray}
N_{\rm e}\,=\, \frac{W_{\rm x}}{\Delta t_{\rm rad}\,p_{\rm s}}
&=&\frac{L_{\rm x}}{p_{\rm s}}\,\approx\,8.5 \cdot 10^{48} ~~{\rm (Model~1)}\\
&=& 4\,\kappa\,\frac{L_{\rm x}}{p_{\rm s}}\,\approx\,2 \cdot 10^{46} ~~{\rm (Model~2)}
\end{eqnarray}
produce the flare. The models require a particle luminosity averaged over the duration 
of the flare of
\begin{eqnarray}
L_{\rm e}\,=\,
\frac{
N_{\rm e}\, \gamma_{\rm 0}\, m_{\rm e}\, c^2} {\Delta t_{\rm rad}}&\approx&
6.3\cdot 10^{45}~{\rm erg~sec}^{-1}~~{\rm (Model~1)}\\
&\approx& 1.5\cdot 10^{43}~{\rm erg~sec}^{-1}~~{\rm (Model~2)}
\label{le}
\end{eqnarray}
For Model 2 it is indeed the observed flare duration and not the 
time $\Delta t_{\rm rad}$ that is  relevant for computing the 
power required for sustaining continued flaring activity.
These luminosities are minimum luminosities to produce the observed X-ray emission.
There may be additional electrons that produce X-ray emission outside the 
{\it BeppoSAX} energy band. 

The electron beam of Model 2 resembles in some aspects the 
blobs filled with emitting particles of the reference model.
This is not too surprising as both models assume that the radiation 
is emitted as synchrotron-Compton emission. 
The different models result in some differences in the phase space distribution
of the electrons and photons. Furthermore, the beams in Models 1 and 2 could 
be made entirely of high-energy particles. The reference model assumes the 
presence of a support medium of rather cold particles, and shocks are 
invoked to explain how the energy of the bulk of cold particles is 
transferred to a few high-energy particles.

It is instructive to compare these jet luminosities to the Eddington luminosity.
Based on bulge stellar dispersion measurements, the best estimate of the black hole mass in Mrk 501 is
$M_{\rm BH}\,\approx$ $10^{9}$ $M_{\odot}$ \citep{Falomo2002,Barth2003}.
The Schwarzschild radius thus is
\begin{equation}
r_{\rm Sch}\,=\,2\,G\,M_{\rm BH}\,/\,c^2\,=\,3\cdot 10^{14}\,\frac{M_{\rm BH}}{10^{9} M_{\odot}}{\rm~cm}
\end{equation}
and the expression for the Eddington luminosity reads
\begin{equation}
L_{\rm Edd}\,=\,
\frac{4\,\pi\,c\,m_{\rm p}\,G\,M_{\rm BH}}{\sigma_{\rm T}}\,=\,
1.25 \cdot 10^{47}~\frac{M_{\rm BH}}{10^9 M_{\odot}}~{\rm erg~sec}^{-1}
\end{equation}
It is roughly 20 and 10$^4$ higher than the minimum luminosities required by 
Models 1 and 2, respectively.
\subsection{Numerical Results}
While the analytic estimates allow us to understand the scaling of the
power-requirements, they are not appropriate for describing the inverse Compton 
component and internal $\gamma$-ray absorption processes as these two depend
on the details of the synchrotron photon energy spectrum.
The following numerical estimates use electron energy spectra $dN_{\rm e}/d\gamma$ instead of
a mono-energetic electron distribution. We use energy spectra that resemble broken power 
laws over small dynamic ranges with a ratio of the maximum to minimum Lorentz 
factor of $\sim100$. The synchrotron power emitted by the leptons at frequency 
$\nu$ per frequency interval d$\nu$ is computed using the standard equation \citep{Rybicki1986}:
\begin{equation}
P_{\rm s}(\nu)\,=\,
c_1\,\int_{\gamma_{\rm min}}^{\gamma_{\rm max}}
d\gamma~\int_0^{\theta_{\rm max}}\sin\theta~d\theta~
\frac{dN_{\rm e}}{d\gamma}~\frac{3\,e^3\,B\,\sin{\theta}}{m_{\rm e}\,c^2}~F(x)
\end{equation}
where the first integral runs over the electron Lorentz factors and
the second averages over the pitch angle distribution.
Here and in the following we use the constant $c_1\,\equiv$ $(1-\cos{\theta_{\rm max}})^{-1}$
to normalize the integrals over the pitch angle distribution.
The function $F(x)$ equals $x\int_x^{\infty}~K\frac{5}{2}(\xi)~d\xi$ with
$K\frac{5}{2}$ the modified Bessel function of $5/2$ order, 
and $x\,=$ $\nu\,/\,\nu_{\rm c}$.
The emitted inverse Compton power is approximately given by
\begin{equation}
P_{\rm ic}(\nu)\,=\,
c_1\,
\int_{\gamma_{\rm min}}^{\gamma_{\rm max}}d\gamma~
\int_{\nu_{\rm min}}^{\nu_{\rm max}}d\nu_{\rm s}~
\int_0^{\theta'_{\rm max}}\sin\theta~d\theta~
\frac{dN_{\rm e}}{d\gamma}~\,c\left(1-\cos{\theta} \right)\,
\sigma_{\rm KN}(y)\,
n_{\rm s}(\nu_{\rm s})\,
\Delta E(y)
\end{equation}
Here we use $\theta'_{\rm max}\,=\,\theta_{\rm max}$ in rough approximation.
The last term in the integrand is the energy that a photon gains 
in a scattering. The other terms give the scattering rate that 
depends on the angle between the electron and photon velocity vectors 
and on the density of synchrotron photons $n_{\rm s}$. 
The Klein-Nishina cross section is to good approximation
\begin{equation}
\sigma_{\rm KN}(y)\,=\,\frac{3}{8}\,\frac{\ln{(2\,y)}+0.5}{y}\,\sigma_{\rm T}
\end{equation}
where $\sigma_{\rm T}$ is the Thomson cross section. The value $y$ is the photon energy 
in the electron rest frame in units of the electron rest mass 
\begin{equation}
y\,=\,
\frac{h\,\nu_{\rm  s}\,\gamma_{\rm  0}\,(1-cos{\theta})}
{m_{\rm e}\,c^2}
\end{equation}
Following \citet{Dermer1993}, we use the approximation
\begin{eqnarray}
\Delta E(y)&=&y\,\gamma\,m_{\rm e}\,c^2 {\rm~for~} y~<~1 {\rm ~and~}\\
&=&\gamma\,m_{\rm e}\,c^2 \hspace*{0.27cm} {\rm~for~} y~\ge~1
\end{eqnarray}
For Model 1, $n_{\rm s}$ can be computed as follows. The total number 
of photons emitted  in a certain frequency interval over the duration
of the flare is $N_{\rm s}(\nu)\,=$ $\Delta t_{\rm obs}\,P_{\rm s}(\nu)\,/$ ${h\,\nu}$.
The emitting particles occupy a volume of $\pi\,R^3$. 
Taking into account that the photon density rises from 0 to its final value
as the emitting particles move through the emission region, the 
time averaged photon density is
\begin{equation}
n_{\rm s}(\nu)\,=\,\frac{1}{2\,\pi\,R^3}\,\frac{\Delta t_{\rm obs}\,P_{\rm s}(\nu)}{h\,\nu}~~{\rm (Model~1)}
\end{equation}
In the case of Model 2, a similar argument gives after some arithmetic:

\begin{eqnarray}
n_{\rm s}(\nu)
&=&\frac{2\,(1-\ln{2})}{\pi\,(\Delta t_{\rm obs}\,c)^3}
\,\frac{\Delta t_{\rm obs}\,P_{\rm s}(\nu)}{h\,\nu}~~{\rm (Model~2)}
\end{eqnarray}
The factor $(1-\ln{2})$ arises from averaging the synchrotron photon density over 
the time the emitting particles travel from $d_1$ to $d_2$, taking into account that
the shell height increases from $H/2$ to $H$ and the number of synchrotron 
photons increases linearly from 0 to its final value.
Finally, the time averaged fluxes received at Earth can be computed from:
\begin{eqnarray}
I(\nu)&=&\frac{P_{\rm s}(\nu)+P_{\rm ic}(\nu)}{\Delta \Omega\, D^2} ~~{\rm (Model~1)}\\
&=&\frac{\Delta t_{\rm rad}\,(P_{\rm s}(\nu)+P_{\rm ic}(\nu))}{\Delta t_{\rm obs}\,\Delta \Omega\, D^2}
\,=\,\frac{P_{\rm s}(\nu)+P_{\rm ic}(\nu)}{4\,\kappa\,\Delta \Omega\, D^2}~~{\rm (Model~2)}
\end{eqnarray}

The optical depth per path length for $\gamma\gamma\,\rightarrow$ $e^+\,e^-$ processes is computed
with the equations of \citet{Gould1967}
\begin{equation}
\frac{d\tau_{\rm int}}{dz}(\nu_{\gamma})
\,=\,c_1\,
\int_0^{\theta_{\rm max}}\sin{\theta}d\theta~
\int_{\nu_{\rm thr}}^{\infty}d\nu~
\sigma_{\gamma\gamma}\,n_{\rm s}(\nu)\,(1-\cos{\theta})
\end{equation}
The first integral runs over the pitch angle distribution and the 
second integrates over the target photon frequencies.
The threshold frequency for pair creation reads
\begin{equation}
\nu_{\rm thr}\,=\,
\frac{2\,(m_{\rm e}\,c^2)^2}{h^2\,\nu_{\gamma}\,(1-\cos{\theta})}
\end{equation}
and the pair-creation cross section is 
\begin{equation}
\sigma_{\gamma\gamma}\,=\,\frac{3\,\sigma_{\rm T}}{16}(1-\beta^2)
\left[2\,\beta\,(\beta^2-2)+(3-\beta^4)\,ln{\left(\frac{1+\beta}{1-\beta}\right)}\right]
~{\rm cm}^2
\label{pc}
\end{equation}
with $\beta\,=$ $\sqrt{1-\nu_{\rm thr}/\nu}$.
In the following, we focus only on absorption by synchrotron photons.
Absorption by external (e.g. disk) photons will briefly
be discussed in Sect.\ 3.2.
For Model 1, the optical depth is approximately $R\,d\tau_{\rm int}/dz$ 
and for Model 2 it is $\frac{1}{2}(d_2-d_1)\,d\tau_{\rm int}/dz$.

The optical depth for extragalactic pair-creation processes is computed based 
on the model of the extragalactic background light (EBL) from \citet{Kneiske2002} 
and \citet{Kneiske2004}. This model is almost identical to the model P045 
of \citet{Aharonian2006} and is thus consistent with the observed 
energy spectra of two distant BL Lac objects detected with the H.E.S.S.\ experiment.
An equation similar to eq.\ (\ref{pc}) is used with the modifications that the 
integral over pitch angles goes from 0 to $\pi$ and the 
EBL target photon density is used.

\begin{figure}[bh]
\begin{center}
\resizebox{8cm}{!}{\plotone{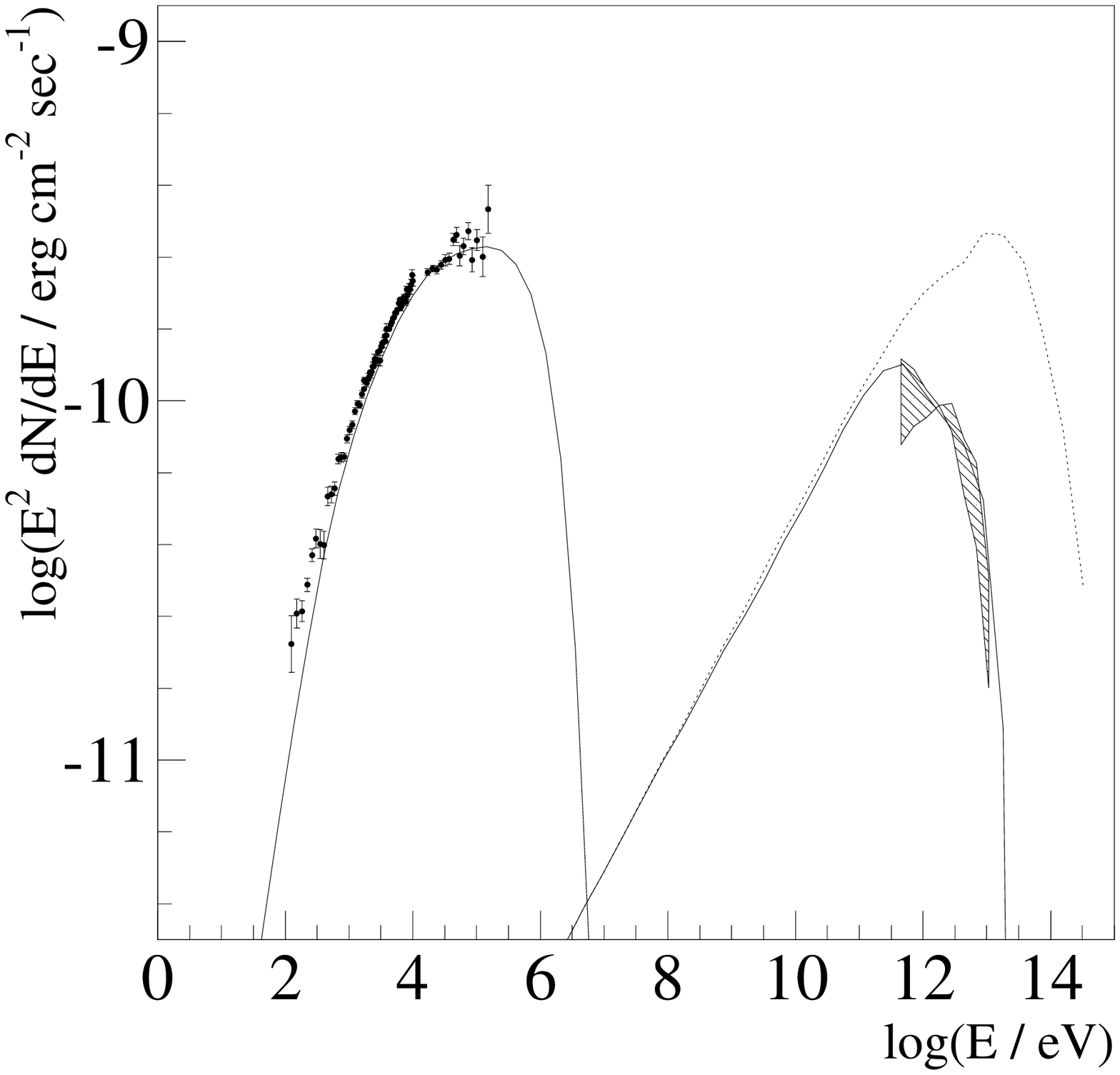}\hspace*{0.2cm}}
\resizebox{8cm}{!}{\plotone{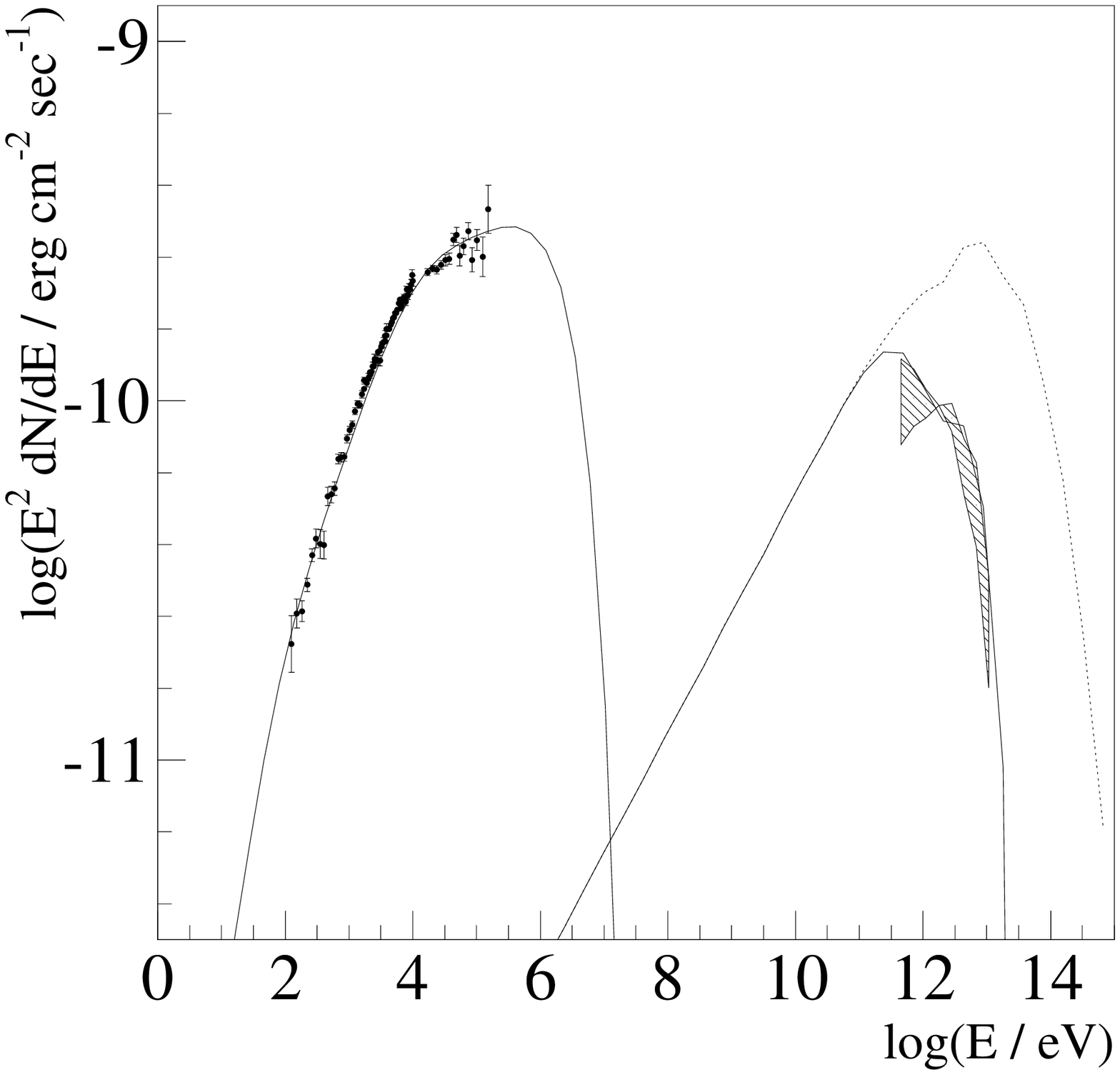}\hspace*{0.2cm}}
\end{center}
\caption{\label{SED}  
Results from modeling the Mrk 501 data from April 16, 1997 with the parallel 
beam model. The left and right panels show fits with Model 1 and Model 2, respectively.
The solid and dotted lines show the emission with and without
accounting for extragalactic absorption in $\gamma\gamma\rightarrow e^+e^-$ pair creation
processes. 
The April 16 flare was the brightest of a large number of flares detected 
during a 6 month period. We show here the {\it BeppoSAX} X-ray and 
{\it CAT} $\gamma$-ray fluxes at only 50\% of their measured values, to take
into account that the observations did not cover the entire flare
and that the fluxes averaged over the entire duration of the flare are 
likely to be lower than those measured at the peak of the flare.
In the case of the {\it CAT} spectrum the shaded area shows the systematic 
uncertainties which dominate over the statistical uncertainties.
The model parameters are given in the text.
}
\end{figure}

The models together with the {\it BeppoSAX} and {\it CAT} energy spectra are shown in Fig.\ \ref{SED}.
We use the {\it BeppoSAX} data reanalyzed by G.\ Fossati (private communication, 2006) and
the {\it CAT} data from \citep{Djannati-Atai1999}. In the case of the {\it CAT} data, we show 
the systematic (rather than statistical) errors which seem to be the dominant ones.
Comparing TeV $\gamma$-ray energy spectra taken with different Cerenkov
telescope experiments (including {\it CAT}) we obtain a one sigma systematic 
error on TeV spectral indices of approximately 0.2. 

The difference of the solid and dotted lines in Fig.\ \ref{SED} shows the effect 
of extragalactic $\gamma$-ray absorption.
Although the emitted inverse Compton components peak at and above $>$5 TeV, the 
absorbed energy spectra are rather soft compared to the measured SEDs.
Models with harder $\gamma$-ray energy spectra can be produced with inverse Compton
processes deeper in the Klein-Nishina regime. We did not implement this 
possibility here as the resulting models do not fully account for the 
$<1$~keV {\it BeppoSAX} data, and would thus require additional X-ray 
emission from downstream plasma to reproduce the $<1$~keV flux.

The specific choices of model parameters result in mechanical luminosities 
of 2.3$\cdot 10^{46}$ erg~sec$^{-1}$ and 2.7$\cdot 10^{44}$ erg~sec$^{-1}$ for 
Model 1 and 2, respectively. 
For Model 1, we fitted the data using $\Delta t_{\rm obs}\,=$ 22~hrs,
$\theta_{\rm max}\,=$ $1^{\circ}$, $\alpha_{\rm max}\,=$ $1^{\circ}$ as
the main input parameters. For Model 2, we used $\Delta t_{\rm obs}\,=$ 12~hrs,
$\theta_{\rm max}\,=$ $2^{\circ}$, and $\alpha_{\rm max}\,=$ $3^{\circ}$.
Compared to the parameters of Model 1, we had to assume for Model~2 a larger solid angle 
into which the radiation is emitted and a shorter $\Delta t_{\rm obs}$ 
in order to reproduce the TeV~$\gamma$-ray flux level.
The power requirement for Model 2 could be reduced to a value close 
to the one in eq.\ (\ref{le}) with smaller $\theta_{\rm max}$ and 
$\alpha_{\rm max}$ values and assuming the presence of external 
seed photons, for example from an outer slower layer of the 
jet similar as in the model of \citet{Ghisellini2005}.

The main conclusions from the more detailed modeling are that (i) 
parallel beam synchrotron self-Compton models are viable 
without requiring external seed photons to account for the 
observed $\gamma$-ray emission, (ii) internal absorption 
effects are small or negligible, and (iii) that the high-energy 
beams require so much power that power efficient beam formation 
models are strongly preferred.
\section{Origin of the parallel electron beam}
\label{BeamFormation}
\subsection{Constraints from the Total Energetics}
\begin{figure}[bh]
\begin{center}
\resizebox{8cm}{!}{\plotone{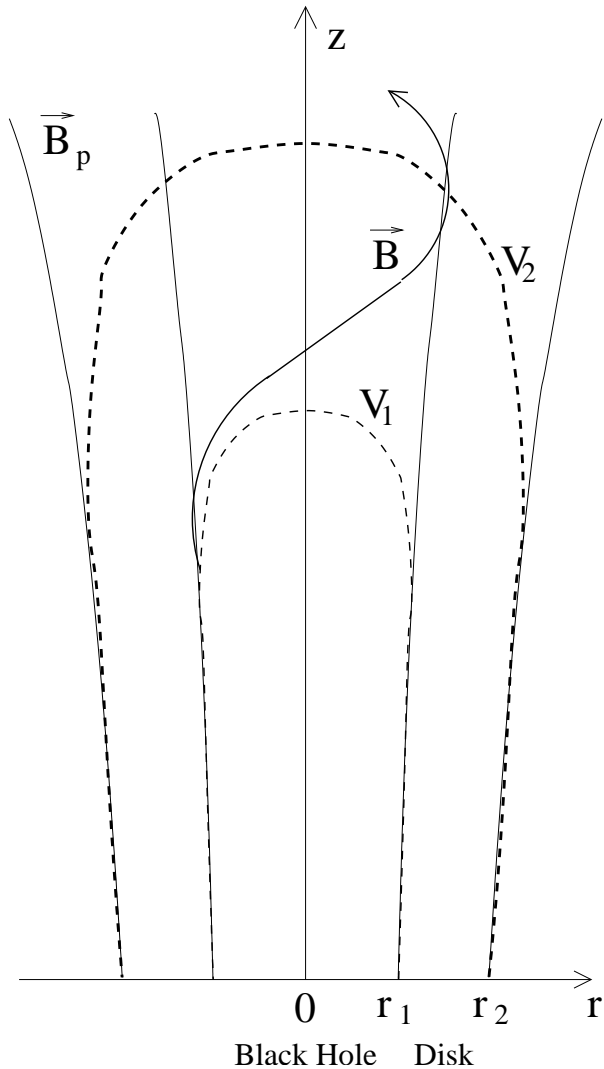}}
\end{center}
\caption{\label{GEOM}  
Snapshot of the magnetosphere and a non-stationary vacuum gap above an accreting
black-hole system. Close to the black hole and the disk, the magnetosphere is filled
with low-energy $e^+/e^-$ pairs that move along electric equipotential 
surfaces (dashed lines). Further away, there is a vacuum gap with a voltage 
drop $V_{12}\,=$ $V_2-V_1$ that can accelerate particles to $>$TeV energies.
See Lovelace, MacAuslan, \& Burns (1979) and Lovelace \& Ruchti (1983)
for a slightly different geometry.}
\end{figure}

We concentrate here on electromagnetic models as they predict powerful large-scale magnetic 
and electric fields that might be able to accelerate particles to very high energies 
and produce large-scale ordered motion. 
Three geometries for accelerating particles in electromagnetic models have 
been discussed in the literature. The models assume that the black hole 
and accretion disk are embedded in a conducting 
magnetosphere. Free charge carriers are created in the magnetosphere 
through cascades involving curvature radiation or inverse Compton scattering and 
pair creation processes. A quasi-stationary configuration is achieved if the charge carriers 
are distributed such that $\vec{E} \cdot \vec{B}\,=0$. 
If this condition is fulfilled, charged particles move along magnetic field lines 
without gaining or loosing energy. The magnetosphere can act as a series of nearly parallel 
conductors with zero resistance along the magnetic field lines that can sustain a 
voltage drop far away from where it was generated. The voltage drop may either be generated 
by the disk itself \citep{Lovelace1976,Blandford1976}, or by the Kerr black hole spinning 
in the horizon-threading magnetic field supported by the disk \citep{Blandford1977,Thorne1986}. 
Katz (2006) argues that the electric field generated by a homopolar generator in a rotating 
magnetized fluid has curls in the observers frame and cannot be shorted out 
at all locations by a stationary charge distribution. It thus seems likely that 
non-stationary vacuum gaps form, which are capable of accelerating particles. 
A possible geometry of the magnetosphere and the gap region 
is shown in Figure~\ref{GEOM}. The formation of the gap close to the rotation axis might explain
the initial collimation of the jet.

In electromagnetic models, the Poynting flux launches the jet and anchors 
the jet to the disk or black hole. For simplicity we focus here on disk models; 
some considerations are applicable to the Blandford-Znajek model as well. 
We assume that the coordinate system $r$, $\phi$, $z$ is aligned with the jet and
the black hole resides at its origin. Directly above the disk, the Poynting flux is
\begin{equation}
\vec{S}\,=\,\frac{c}{4\,\pi}\,\vec{E}\times \vec{B}\,=\,\frac{c}{4\,\pi}\,\vec{E}_{\rm p}\times \vec{B}_{\rm t}
\label{pf}
\end{equation}
where the subscripts p and t denote the poloidal ($r$ and $z$) 
and toroidal ($\phi$) components, respectively.  The second equality follows 
from the fact that the toroidal electric field vanishes for a stationary 
axisymmetric solution. 
The condition that the electric field $\vec{E}'$ in the frame co-rotating 
with the disk material vanishes gives
\begin{equation}
0\,=\,\vec{E'}\,=\,\vec{E}_{\rm p}\,+\frac{1}{c}\left(\vec{\Omega}\times \vec{r}\right)\times \vec{B}_{\rm p}
\end{equation}
with $\vec{\Omega}\,=$ (0,0,$\Omega$) the angular velocity of the disk material 
at location $\vec{r}$. Thus, the poloidal electric field satisfies
\begin{equation}
\vec{E}_{\rm p}\,=\,-\frac{1}{c}\,\left(\vec{\Omega}\times \vec{r}\right)\times \vec{B}_{\rm p}
\label{ep}
\end{equation}
Combining eqs.\ (\ref{pf}) and (\ref{ep}), the magnitude of the Poynting flux is
\begin{equation}
S\,=\,\frac{1}{4\pi}\,\Omega\,r\,B_{\rm p}\,B_{\rm t}
\label{pf2}
\end{equation}
If we now use
\begin{equation}
\Omega\,r\,=\,\left(\frac{G\,M_{\rm BH}}{r}\right)^{1/2}
\end{equation}
and assume
\begin{eqnarray}
B_{\rm t}(r)&=&B_{\rm t,r_1} \, (r/r_1)^{-1}\\
B_{\rm p}(r)&=&B_{\rm p,r_1} \, (r/r_1)^{-1}
\end{eqnarray}
the power transported by the Poynting flux perpendicular to the disk surface is
\begin{eqnarray}
L_{\rm disk}
\,=\,2\,\pi\,\int_{r_1}^{r_2}\,S\,r\,dr
\,=\, \frac{6^{3/2}}{4\,c^3}\,\left(G\,M_{\rm BH}\right)^2\,B_{\rm t}\,B_{\rm p}
\,\approx\, 5\cdot 10^{44}\,\frac{M_{\rm BH}}{10^9\,M_{\odot}}\,\frac{B_{\rm t,r_1}}{10^3\,\rm G}\,\frac{B_{\rm p,r_1}}{200\,\rm G}
{\rm ~erg~sec}^{-1}
\label{ldisk}
\end{eqnarray}
where we used $r_1\,\approx\,3\,R_{\rm Sch}$ and $r_2\,\gg\,r_1$.
Thus, magnetic fields between $10^2$ and $10^3$~G are needed 
to form the beam of Model~2. Here and in the following, we use the
rather high power requirement from the numerical SSC calculations.
The reader should keep in mind that external Compton models would require
$\sim$20 times less power. For Model 1, $B_{\rm t}$ and $B_{\rm p}$ have 
to be both roughly ten times higher. 
Various estimates of the magnetic field in 
accretion disks have been discussed by \cite{Ghosh1997}.
Magnetic fields of $\simeq 10^4$~G seem likely from 
dimensional arguments and numerical simulations.

The voltage drop across the disk from $r_1$ to $r_2\,\gg\,r_1$ is 
\begin{equation}
V_{\rm 12}\,=\,\int_{r_1}^{r_2}\frac{1}{c}\,r\,\Omega\,B_{\rm p}\, dr\,=\,
\frac{\sqrt{24}\,G\,M_{\rm BH}\,B_{\rm p,r_1}}{c^2}\,\approx\,
4.3\cdot 10^{19}\,\frac{M_{\rm BH}}{10^9\,M_{\odot}}\,\frac{B_{\rm p,r_1}}{200\,\rm G}{\rm~V}
\label{V12}
\end{equation}
If $r_2\,=$ 2$\,r_1$ (rather than $r_2\gg r_1$), $L_{\rm disk}$ and $V_{\rm 12}$
are smaller by a factor of $(1-1/\sqrt{2})$.
A natural mechanism for explaining the variable nature of the X-ray and TeV gamma-ray emission
from blazars is screening of the voltage $V_{12}$ by electron/positron pairs.
The Blandford-Znajek mechanism results in a qualitatively similar relations 
between the magnetic field, the electric fields, and the emitted power.
The main difference is that the EMF is generated close to the event horizon 
rather than at the inner region of the accretion disk.

Given a poloidal magnetic field with $B_{\rm p,r_1}\,\approx$ 200~G at the base of the jet,
we can estimate the magnetic field at the distance $d_2$ in the simple case that
the poloidal magnetic field scales inversely proportional to the radius of the 
emission zone squared. For Model 2 we obtain 
\begin{equation}
B_{\rm d1}\,=\,B_{\rm r_1}\,\left(\frac{r_1}{H}\right)^2\,=\,
0.013~\left(\frac{B_{r_1}}{\rm 200~G}\right)~
\left(\frac{H}{3.7\cdot 10^{16}\rm~cm}\right)^{-2}
{\rm~G}
\end{equation}
where $H$ is the radius of the emission zone perpendicular to 
the jet axis. This simple estimate agrees well with the value inferred 
from modeling the data (see eq.\ (\ref{B})). 
\subsection{Beam Formation}
The number of electrons required to explain the X-ray and TeV $\gamma$-ray emission is so 
large that charge neutrality of the beam is important as otherwise the jet would expand
too rapidly in the direction perpendicular to the jet axis. The maximum energy to which particles 
are accelerated depend on the relative magnitude of the energy gain and energy loss rates.
The energy gain rate depends on $V_{12}$ and the distance over which the voltage drop occurs.
Energy loss mechanisms are curvature radiation, synchrotron emission and inverse Compton emission.
With curvature radii on the order of $r_1$, curvature losses are only important at very high $\gg$TeV energies.
The magnitudes of the synchrotron and inverse Compton losses are highly uncertain as they
depend on the magnetic field strength and particle pitch angle distribution, and on the 
intensity of the ambient photon field in the acceleration region, respectively. Synchrotron 
losses are negligible if electrons drift along magnetic field lines.
Assuming scattering in the Thomson regime, a 20~TeV electrons loses the energy 
\begin{equation}
\Delta E_{\rm ic}\,=\,\gamma_0\next^2\,(1-\cos{\vartheta})^2\,l_{\rm acc}\,\sigma_{\rm T}\,u_{\rm a}
\end{equation}
when traveling a distance $l_{\rm acc}$ through an ambient photon field with energy
density $u_{\rm a}$. Here, $\vartheta$ is the angle between the electron 
and photon velocity vectors. The energy losses are negligible when $u_{\rm a}$ satisfies
\begin{eqnarray}
u_{\rm a}&\ll &
\frac{m_{\rm e}\,c^2}{\gamma_0\,(1-\cos{\vartheta})^2\,l_{\rm acc}\,\sigma_{\rm T}} \,\approx\, 0.002
{\rm~erg~cm}^{-3}
\end{eqnarray}
for $\vartheta\,=$ 30$^{\circ}$ and $l_{\rm acc}\,=\,r_1$.
The ambient photon energy density corresponds to a luminosity of
\begin{equation}
L_{\rm a}\,=\,4\pi\,(30\,r_1)^2\,c\,u_{\rm a}\,\approx\,5\cdot 10^{41}
{\rm erg~sec}^{-1}
\end{equation}
if the particle acceleration region is at a distance $30\,r_1$ from the photon source,
presumably the accretion disk. Comparing the minimum beam luminosity 
required for producing the observed X-ray and TeV $\gamma$-ray emission 
(eq.\ (\ref{le})) with $L_{\rm a}$, we see that the model requires an 
accretion flow with a radiative efficiency of 3\% or less.
Higher disk luminosities are possible if the emission frequency is 
sufficiently high that the high-energy electrons interact only 
in the Klein-Nishina regime.

%
%
\citet{Lovelace1976} assumed that protons may be accelerated all the way to 
ultra high energies and that quasars thus may be accelerators of ultra high 
energy cosmic rays (see \citealt*{Boldt1999,Boldt2000,Levinson2000} 
for similar recent papers). He stipulated that the flow of high-energy protons 
may entrain or pick up electrons. Assuming that the high-energy electrons and 
protons would move with identical velocities and Lorentz factors, the 
acceleration of protons would increase the minimum beam luminosity by the 
proton to electron mass ratio. For Model 1, the required luminosity would 
exceed the Eddington luminosity by at least two orders of magnitude.
For Model 2, accretion with a few times the Eddington rate would 
be sufficient. With $B_{\rm p,r_1}\approx$0.17~G, 
eq.\ (\ref{V12}) predicts proton energies of 
$\approx 3\cdot 10^{16}$~eV and just the right electron energy 
of $\approx$20~TeV. However, a prohibitively strong toroidal magnetic field 
exceeding $10^6$~G would be required so that the Poynting flux 
can power the massive electron-proton beam.

The acceleration of electrons or positrons with subsequent entrainment of 
oppositely charged leptons would result in a beam with a much lower power.
However, eq.\ (\ref{V12}) predicts an adequate voltage drop for a very weak 
poloidal magnetic field with $B_{\rm p,r_1}\,\approx$ $2 \cdot 10^{-4}$ G. 
Even for Model 2, the beam power requires again a toroidal 
magnetic field exceeding $10^6$~G.
Stronger $B_{\rm p,r_1}$ and weaker $B_{\rm t,r_1}$ would be viable
if the leptons are accelerated to energies exceeding 40~TeV, and then entrain
both electrons and positrons, slowing them down to a mean energy of 
$\sim$20~TeV per lepton.

Acceleration of electrons or positrons with $B_{\rm t}\sim B_{\rm p}$
would produce a few particles with very high energies. A natural way of 
transferring the energy from a few high-energy particles to many low-energy 
particles are cascades. Electromagnetic cascades in AGN jets have been discussed 
by \citep{Burns1982,Blandford1995,Levinson1995}. 
A generic discussion of electromagnetic cascades 
in the $>$TeV regime has been given in \citep{Svensson1987}. 
Unfortunately, cascade models need considerable fine tuning to produce 
the electron beams with the right properties. 
We briefly go through a specific scenario to emphasize some of the 
relevant difficulties. 
Electrons are accelerated until the energy gains  in the electric 
field $E_{12}\,=V_{12}/l_{\rm acc}$ along magnetic field lines equal the energy losses. 
If curvature losses dominate, the energy loss rate reads
\begin{equation}
p_{\rm c}\,=\,\frac{2}{3}\,\frac{e^2\,c\,\gamma^4}{\rho^2}
\end{equation}
with $\rho$ an average curvature radius, which gives a maximum 
Lorentz factor of
\begin{equation}
\gamma_{\rm max}\,=\,\left(\frac{3}{2}\,\frac{(V_{12}/l_{\rm acc})\,\rho^2}{e}\right)^{1/4}\,\approx\,
10^{11}
\left(\frac{V_{\rm 12}}{10^{19}\rm~V}\right)^{1/4}\,
\left(\frac{l_{\rm acc}}{0.15\,r_1}\right)^{-1/4}\,
\left(\frac{\rho}{r_1}\right)^{1/2} 
\end{equation}
which corresponds to an energy of 50~PeV.
The highest-energy electrons emit curvature photons at frequency
\begin{equation}
\nu_{\rm c}\,=\,\frac{3}{4\pi}\,\frac{\gamma_{\rm max}\next^3\,c}{\rho}\,\approx
10^{28}
\left(\frac{V_{\rm 12}}{10^{19}\rm~V}\right)^{3/4}\,
\left(\frac{l_{\rm acc}}{0.15\,r_1}\right)^{-3/4}\,
\left(\frac{\rho}{r_1}\right)^{-1/2} 
{\rm Hz}
\end{equation}
corresponding to a photon energy of $\simeq$40~TeV.
Overall, each lepton drifting through the acceleration region 
emits $\gamma$-rays with a total energy of 
$e\,V_{12}-\gamma_{\rm max}\,m_{\rm e}\,c^2$, while it escapes with 
a relatively small amount of 
kinetic energy $\sim\gamma_{\rm max}\,m_{\rm e}\,c^2$.
A part of the energy in the $\gamma$-ray beam may be converted back to 
the leptonic sector if the optical depth for pair creation is approximately 
unity. The pair creation cross section (eq.\ (\ref{pc})) reaches its 
maximum value of $\approx 1/4\,\sigma_{\rm T}$ for target photons of 
frequency $\nu_{\rm a}\approx 2\,\nu_{\rm thr}$.
Assuming that the target photons have the frequency $\nu\,\approx $ $2\nu_{\rm thr}$, 
and that the curvature $\gamma$-rays are emitted at a distance
$30\,r_1$ from the target photon source, we infer a minimum target 
photon luminosity of 1.4$\cdot 10^{41}$ 
erg sec$^{-1}$ for which the pair creation optical depth for 
40 TeV $\gamma$-rays escaping to infinity equals unity.
The model  requires fine-tuning as seed photons of frequency 
$\approx 2\,\nu_{\rm thr}$ are needed to suppress inverse Compton 
processes of $>$20~TeV electrons.
If the target photon field is sufficiently intense, a cascade with several
generations of photons and pairs can be initiated. The end-product of the 
cascade then depends on the energy spectrum and spatial gradient of 
the target photons. 

Proton induced synchrotron/pair-creation cascades (PIC) 
in blazars have been discussed by \citep{Mannheim1993,Muecke2001,Muecke2003}.
These models produce $\gamma$-rays as synchrotron emission of 
high-energy electrons/positrons with possible contributions from other secondary
cascade particles. The models require not only high-energy protons, but also
co-spatially accelerated electrons. The latter emit the 
radiation field that causes the protons to photo-produce mesons, and
explain the observed X-ray emission. PIC models had originally been 
proposed in the framework of the shock-acceleration picture.
A re-evaluation of these models with regards to accelerating the protons 
and electrons in the strong electric fields in the surrounding of a 
black hole may be a worthwhile enterprise, but
is outside of the scope of this paper.
\section{Discussion}
\label{discussion}
This paper discusses the possibility that the X-ray and TeV $\gamma$-ray 
radiation from BL Lac type objects is emitted by parallel 
electron-positron beams that are accelerated by strong electric fields
close to the accreting central black hole. For the first time we attempt 
to explain both the X-ray and TeV $\gamma$-ray emission from these objects 
with such a model. Fitting the model to data from the BL Lac object Mrk 501, we 
find that the particle acceleration zone and the X-ray and TeV $\gamma$-ray emission zone
have to be spatially distinct. Otherwise, the large magnetic fields required to 
launch the jets prohibit the simultaneous emission of the X-rays and TeV $\gamma$-rays
as synchrotron and inverse Compton radiation, respectively.
Previous papers discussing the high-energy emission from blazars dismissed the
possibility that the high-energy particles producing the observed radiation 
may be accelerated close to the black hole. It was believed that the 
co-spatially emitted radio to optical emission would cause an inverse 
Compton catastrophe, causing the high-energy particles to lose all 
their energy. For several reasons this argument is not valid 
for all blazars. First, in contrast to the situation in powerful quasars,
the accretion disks of BL Lac objects are not radiating efficiently.
For most BL Lac objects, there are only upper limits on the 
disk luminosity. Furthermore, a common assumption had been that the 
observed radio to optical radiation from BL Lac objects
was emitted co-spatially with the X-ray and TeV $\gamma$-ray emission.
For some objects like Mrk 501, this assumption had to be dropped even for 
the reference model, as the upper limits on the size of the emission 
region from the observed flux variability time scales resulted in 
synchrotron self-absorption cut-offs in the 100-1000~GHz range. 
Thus, even the standard model requires that at least the radio emission 
is produced further downstream in the jet than the X-ray and TeV $\gamma$-ray emission.
In BL Lac objects, the observation campaigns carried through so far have not
produced solid evidence for a correlation between the X-ray/TeV gamma-ray fluxes 
and the infrared-optical fluxes. Thus, also the infrared and optical emission may be
produced downstream of the X-ray and TeV $\gamma$-ray emission, or even by independent
processes. Two other factors contribute to the suppression of an inverse Compton catastrophe.
In the parallel beam model the emitting particles and the photons travel 
almost in parallel, greatly reducing their interaction rate;
furthermore, inverse Compton interactions of $>$TeV electrons and positrons 
with all photons with wavelengths shorter than a few microns are Klein-Nishina suppressed.

In the reference model, the jet is launched electromagnetically and the 
jet power is subsequently transferred to particles which carry it to
large distances from the accretion system. Shocks in the jet medium 
transfer the energy transported by a large number of cold particles 
to high-energy particles that emit the observed radiation. 
Compared to this reference model, the model discussed here avoids 
the need for transferring the power first from Poynting flux to a 
large number of particles and then to transfer it back to a few high-energy 
particles. Furthermore, the parallel electron-positron beam model can cope with 
some of the difficulties of the reference model:
\begin{enumerate}
\item The model can explain the ``odd'' energy spectra of emitting 
electrons/positrons inferred from synchrotron-Compton fits to BL Lac 
data with more ease than the Fermi acceleration mechanisms. 
The most striking features are particle distribution functions almost
resembling ``delta-functions'' or Maxwellian distributions.
Such particle distribution might result from the acceleration and/or
cascading processes discussed above.
\item The parallel beam model accounts for the non-detection of one of the
tell-tale signatures of Fermi-type acceleration mechanisms, namely a particularly soft
energy spectrum during the early rising phases of flares.
\item A consequence of the nearly parallel flow of the emitting particles
and photons in the emission region are nearly simultaneous variations of the synchrotron 
and inverse Compton fluxes if fluxes emitted by electrons of similar energies are 
sampled, and if complications arising from cooling of electrons and thus an 
increase in seed photons can be neglected. 
Synchrotron self-Compton models in which the emitting particles move isotropically in the 
jet frame, predict that the synchrotron fluxes should rise faster than the inverse 
Compton fluxes \citep{Coppi1999}.
Although such a time lag has long been searched for, it has eluded detection so far.
\item Sources like Mrk 501 and Mrk 421 show extended flaring periods with many flares.
Furthermore, observations of Mrk 501 in 1997 showed an astonishing stability of the TeV $\gamma$-ray 
energy spectrum during the entire observation campaign \citep{Krawczynski2000}. 
The X-ray and TeV $\gamma$-ray fluxes followed the same correlation during many distinct 
flares over a time period of several months. 
In the parallel beam model, regular and rather uniform flaring might result from alternating 
between shorting out and evacuating the particle acceleration region.
Particle-accelerating recollimation shocks at certain typical distances from 
the central engines 
provide a viable alternative explanation.
In contrast, models in which flares are produced by collisions of plasma blobs 
have difficulties to do so \citep{Tanihata2003}.
\item In the reference model, the high-energy particles move
isotropically. As the particle pressure dominates over the 
magnetic field pressure by many orders of magnitude
\citep{Kino2002,Krawczynski2002}, the model has difficulties
to explain why the particles do not flow out of the emission 
volume with the speed of light. In the parallel beam model, the 
motion of the emitting particles along ordered magnetic field lines
explains the beam collimation more naturally.
\end{enumerate}
The parallel beam model has its own challenges. The discussion in the previous 
section indicates that direct particle acceleration followed by entrainment 
of ambient matter and the study of electromagnetic cascades in the TeV-PeV 
energy regime are areas where more detailed modeling is required.
Another area for future work is more detailed time resolved simulations 
of the temporal evolution of the beam.

In the introduction we mentioned the discrepancy between jet bulk 
Lorentz factors inferred from the reference model to the X-ray and
TeV $\gamma$-ray data, and from VLBA observations.
Both, the reference model, and the parallel beam model can
solve this problem by positing that the emission of the X-rays and 
TeV $\gamma$-rays is the start of a drastic energy dissipation 
of the jet. Model 2 is radiatively very efficient so that 
a considerable fraction of the jet energy goes directly 
into radiation. If radiatively inefficient models apply, the
jet may slow down by entraining ambient material.

In the reference model, the difference between the emission of quasars and 
BL Lac objects is commonly attributed to a difference of the maximum 
energy of accelerated particles owing to more efficient inverse Compton cooling
of accelerated particles in the intense radiation fields of 
quasars \citep{Ghisellini1998}.
The parallel beam model has the potential to explain not only that
the SEDs of quasars and BL Lac objects are different, but also that
the kpc-scale jets are different.
In the reference model, the intense photon fields only affect the particles 
accelerated by the jet far away from the central engine. In the 
parallel beam model, they can affect the process of jet 
formation itself. Thus, the intense radiation fields of quasars may 
prevent a parallel particle beam from forming altogether, and may 
allow a different jet formation mechanism to produce the 
powerful quasar jets.
\acknowledgements
HK thanks Jonathan Katz, Peter M\'esz\'aros, and Charles Dermer 
for comments. He acknolwedges enlightening discussions with J.\ Buckley, 
P.\ S.\ Coppi, and I.\ V.\ Jung. HK is indebthed to G.~Fossati 
and I.~Jung for sharing the re-analyzed {\it BeppoSAX} data. 
HK gratefully acknowledges support by the DOE through the 
Outstanding Junior Investigator program.

\clearpage

\end{document}